%% file: parityPaper.tex
\documentclass[aps,prl,twocolumn,superscriptaddress,altaffilletter,nobibnotes,nofootinbib,10pt,showpacs,showkeys,preprintnumbers]{revtex4-1}

\usepackage[normalem]{ulem}
\usepackage{graphicx}
\usepackage{dcolumn}
\usepackage{bm}
\usepackage{lineno}
\usepackage{float}
\usepackage{color}
\definecolor{dgreen}{cmyk}{1.,0.,1.,0.2}        
\definecolor{orange}{cmyk}{0.,0.353,1.,0.}    

\usepackage{preprintcover}
\PreprintIdNumber{CERN-PH-EP-2012-183}
\PreprintCoverPaperTitle{Charge separation relative to the reaction plane\\in Pb--Pb collisions at $\sqrt{s_{\rm NN}}= 2.76$~TeV}
\PreprintCoverAbstract{
Measurements of charge dependent azimuthal correlations with the ALICE detector at the 
LHC are reported for Pb--Pb collisions at $\sqrt{s_{\rm NN}} = 2.76$~TeV. Two-- and three--particle 
charge--dependent azimuthal correlations in the pseudo--rapidity range $\left|\eta \right| < 0.8$ 
are presented as a function of the collision centrality, particle separation in pseudo--rapidity, 
and transverse momentum. A clear signal compatible with a charge--dependent 
separation relative to the reaction plane is observed, which shows little or no collision energy 
dependence when compared to measurements at RHIC energies. This provides a new insight for understanding the nature of the charge dependent azimuthal correlations observed at RHIC and LHC energies.
}
\PreprintJournalName{PRL}

\newcommand{ \phia }{\varphi_{\alpha}}
\newcommand{ \phib }{\varphi_{\beta}}
\newcommand{ \mean }[1]{\left\langle #1 \right\rangle}   
\newcommand{\pT}{$p_{\rm{T}}$~}

\newcommand{\p}{$P$}
\newcommand{\cp}{$CP$}

\newcommand{\SPD}{\rm{SPD}}

\newcommand{\TPC}{\rm{TPC}}

\newcommand{\VZERO}{\rm{VZERO}}
\newcommand{\ZDC}{\rm{ZDC}}
\newcommand{\cme}{\rm{Chiral Magnetic Effect}}

\newcounter{vers}

\begin{document}
\setcounter{page}{2}


\setlength{\linenumbersep}{6pt}

\title{Charge separation relative to the reaction plane in Pb--Pb collisions at $\sqrt{s_{\rm NN}}= 2.76$~TeV}

\collaboration{ALICE Collaboration}
\input{authorListAPS.tex}


\begin{abstract} 
Measurements of charge dependent azimuthal correlations with the ALICE detector at the 
LHC are reported for Pb--Pb collisions at $\sqrt{s_{\rm NN}} = 2.76$~TeV. Two-- and three--particle 
charge--dependent azimuthal correlations in the pseudo--rapidity range $\left|\eta \right| < 0.8$ 
are presented as a function of the collision centrality, particle separation in pseudo--rapidity, 
and transverse momentum. A clear signal compatible with a charge--dependent 
separation relative to the reaction plane is observed, which shows little or no collision energy 
dependence when compared to measurements at RHIC energies. This provides a new insight for understanding the nature of the charge dependent azimuthal correlations observed at RHIC and LHC energies.
\end{abstract}
\maketitle


\input{parityPaperMain.tex}

\section{Acknowledgements}
\input{acknowledgements.tex}

\input{references.tex}

\end{document}

%% file: authorListAPS.tex
%
\author{B.~Abelev}
\affiliation{Lawrence Livermore National Laboratory, Livermore, California, United States}
\author{J.~Adam}
\affiliation{Faculty of Nuclear Sciences and Physical Engineering, Czech Technical University in Prague, Prague, Czech Republic}
\author{D.~Adamov\'{a}}
\affiliation{Nuclear Physics Institute, Academy of Sciences of the Czech Republic, \v{R}e\v{z} u Prahy, Czech Republic}
\author{A.M.~Adare}
\affiliation{Yale University, New Haven, Connecticut, United States}
\author{M.M.~Aggarwal}
\affiliation{Physics Department, Panjab University, Chandigarh, India}
\author{G.~Aglieri~Rinella}
\affiliation{European Organization for Nuclear Research (CERN), Geneva, Switzerland}
\author{A.G.~Agocs}
\affiliation{KFKI Research Institute for Particle and Nuclear Physics, Hungarian Academy of Sciences, Budapest, Hungary}
\author{A.~Agostinelli}
\affiliation{Dipartimento di Fisica dell'Universit\`{a} and Sezione INFN, Bologna, Italy}
\author{S.~Aguilar~Salazar}
\affiliation{Instituto de F\'{\i}sica, Universidad Nacional Aut\'{o}noma de M\'{e}xico, Mexico City, Mexico}
\author{Z.~Ahammed}
\affiliation{Variable Energy Cyclotron Centre, Kolkata, India}
% ...

%% file: parityPaperMain.tex
The possibility to observe parity violation in the strong interaction using relativistic heavy--ion 
collisions has been discussed for many years \cite{Ref:TDLee,Ref:Morley,Ref:Kharzeev}. 
In quantum chromodynamics (QCD), this symmetry violation originates in the interaction 
between quarks and topologically non--trivial gluonic fields, instantons, and sphalerons 
\cite{Ref:InstantonsSphalerons}. This interaction, which is characterised by the topological 
charge \cite{Ref:ChernSimons}, breaks the balance between the number of quarks with 
different chirality, resulting in a violation of the \p--~and \cp--symmetry. 
In \cite{Ref:BField1,Ref:TheoryExpectations}, it was suggested that in the vicinity of the deconfinement phase transition, and under the influence of 
the strong magnetic field generated by the colliding nuclei,the quark spin alignment along the direction of the angular momentum (i.e. the direction of the magnetic field) and the imbalance of the left-- and right--handed quarks, generates an electromagnetic current. The experimental search of these effects
has intensified recently, following the realisation that the consequent quark fragmentation into charged hadrons results in a charge separation along 
the direction of this magnetic field, and perpendicular to the reaction plane (the plane of symmetry 
of a collision defined by the impact parameter vector and the beam direction).
This phenomenon is called the \cme~(CME). Due to fluctuations in the sign of the topological charge, 
the resulting charge separation averaged over many collisions is zero. This makes the observation 
of the CME possible only via \p--even observables, expressed in terms of two-- and multi--particle 
correlations. The previous measurement of charge separation by the STAR Collaboration~\cite{Ref:STAR} 
is consistent with the qualitative expectations for the CME, and has triggered an intense 
discussion~\cite{Ref:Koch,Ref:Muller,Ref:Pratt,Ref:Zhitnitsky,Ref:Toneev}. 

A large source of uncertainty in the theoretical consideration of the CME is related to the expected centre--of--mass 
energy dependence. In \cite{Ref:TheoryExpectations}, the authors argued that  the uncertainty in making any 
quantitative prediction relies on the time integration over which the magnetic field develops and decays. As long as 
a deconfined state of matter is formed in a heavy--ion collision, the magnitude of the effect should either not change 
or should decrease with increasing energy \cite{Ref:TheoryExpectations}. In addition, in \cite{Ref:Zhitnitsky} it is also 
suggested that there should be no energy dependence between the top RHIC and the LHC energies, based on 
arguments related to the universality of the underlying physical process, without however explicitly quantifying what 
the contribution of the different values and time evolution of the magnetic field for different energies will  be. On the 
other hand, in \cite{Ref:Toneev} it is argued that the CME should strongly decrease at higher energies, because the 
magnetic field decays more rapidly. Such spread in the theoretical expectations makes it important to measure the 
charge dependent azimuthal correlations at the LHC, where the collision energy is an order of magnitude higher 
compared to RHIC.

In this Letter we report the measurement of the charge--dependent azimuthal correlations at mid--rapidity in Pb--Pb 
collisions at the centre of mass energy per nucleon pair $\sqrt{s_{NN}} = 2.76$~TeV by the ALICE Collaboration at 
the LHC.

\vspace{-0.08cm}
Azimuthal correlations among particles produced in a heavy--ion collision provide a powerful tool 
for the experimental study of particle production with respect to the reaction plane, which is usually 
quantified by the anisotropic flow coefficients, v$_n$, in a Fourier decomposition \cite{Ref:FlowHarmonics}. 
Local violation of parity symmetry results in the additional P--odd sinus terms \cite{Ref:Kharzeev,Ref:Sergey3particleCorrelator,Ref:STAR}:

\begin{eqnarray}
\frac{dN}{d\varphi_{\alpha}} \sim 1 + 2\sum_n \left[\mathsf{v}_{n,\alpha} \cos(n \Delta \varphi_{\alpha}) + \mathrm{a}_{n,\alpha} \sin(n \Delta \varphi_{\alpha})\right],
\label{Eq:Fourier}
\end{eqnarray}

\noindent where $\Delta \varphi_{\alpha}=\varphi_\alpha - \Psi_{\rm RP}$ is the azimuthal angle 
$\varphi_\alpha$ of the charged particle of type $\alpha$ relative to the reaction plane angle, $\Psi_{\rm RP}$. 
The leading order coefficient $\mathrm{a}_{1,\alpha}$ reflects the magnitude while the higher orders 
($\mathrm{a}_{n,\alpha}$ for $n > 1$) describe the specific shape in azimuth of the effects from local 
parity violation. We thus employ a multi--particle correlator \cite{Ref:Sergey3particleCorrelator} 
which probes the magnitude of the $\mathrm{a}_1$ coefficient, and at the same time suppresses the 
background correlations unrelated to the reaction plane:

\begin{eqnarray}
\langle \cos(\varphi_{\alpha} + \varphi_{\beta} - 2\Psi_{\rm RP}) \rangle &=&
\mean{\cos\Delta \phia\, \cos\Delta \phib}
\nonumber \\
&-&\mean{\sin\Delta \phia\,\sin\Delta \phib}.
\label{Eq:3ParticleCorrelator}
\end{eqnarray}
The indices $\alpha$ and $\beta$ refer to the charge of the particles. The brackets denote an average over 
the particle pairs within the event as well as an average over the analysed events. In practice, the reaction 
plane angle is not known and is estimated by constructing the event plane using azimuthal particle distributions. 
In Eq.~\ref{Eq:3ParticleCorrelator}, the terms $\mean{\cos\Delta \phia\, \cos\Delta \phib}$ and $\mean{\sin\Delta \phia\,\sin\Delta \phib}$ quantify the correlations in-- and out--of plane, respectively. The latter one is sensitive to the charge 
correlations resulting from the CME: $\mean{\sin\Delta \phia\,\sin\Delta \phib} \sim \langle a_{1,\alpha} a_{1,\beta} \rangle$. 
The construction of the correlator in Eq.~\ref{Eq:3ParticleCorrelator} as the difference between these two contributions 
suppresses correlations not related to the reaction plane orientation (non--flow). The contribution from the CME to the 
correlations of pairs of particles with same and opposite charge is expected to be similar in magnitude and opposite in 
sign. This expectation could be further modified by the medium created in a heavy--ion collision, that may result in the 
dilution of the correlations between particles with opposite sign \cite{Ref:BField1,Ref:TheoryExpectations}. In order to 
evaluate each of the two terms in Eq.~\ref{Eq:3ParticleCorrelator}, we also measure the two particle correlator:
%
\begin{eqnarray}
\langle \cos(\varphi_{\alpha} - \varphi_{\beta}) \rangle 
&=&
\mean{\cos\Delta \phia\, \cos\Delta \phib} 
\nonumber \\
&+&
\mean{\sin\Delta \phia\,\sin\Delta \phib},
\label{Eq:2ParticleCorrelator}
\end{eqnarray}
which in contrast to the correlator in Eq.~\ref{Eq:3ParticleCorrelator} is independent of the reaction 
plane angle and susceptible to the large $P$--even background contributions. The combination of 
these correlators provides access to both components, $\mean{\cos\Delta \phia\, \cos\Delta \phib}$ 
and $\mean{\sin\Delta \phia\, \cos\Delta \phib}$, which is important for detailed comparisons with 
model calculations.

It should be pointed out that both correlators of Eq.~\ref{Eq:3ParticleCorrelator} and Eq.~\ref{Eq:2ParticleCorrelator} 
could be affected by background sources. In \cite{Ref:Pratt}, it is argued that the effect of momentum conservation 
influences in a similar way the pairs of particles with opposite and same charge, and could result into a potentially 
significant correction to both $\langle \cos(\varphi_{\alpha} + \varphi_{\beta} - 2\Psi_{\rm RP}) \rangle$ and 
$\langle \cos(\varphi_{\alpha} - \varphi_{\beta}) \rangle$. Also in \cite{Ref:Pratt}, it was suggested that local charge 
conservation effects may be responsible for a significant part of the observed charge dependence of the correlator
$\langle \cos(\varphi_{\alpha} + \varphi_{\beta} - 2\Psi_{\rm RP}) \rangle$. Recent calculations \cite{Ref:TeaneyYan} 
suggest that that the correlator in Eq.~\ref{Eq:3ParticleCorrelator} may have a negative (i.e. out--of--plane), charge 
independent, dipole flow contribution originating from fluctuations in the initial energy density of a heavy--ion collision. 


A description of the ALICE detector and its performance can be found in \cite{Ref:ALICE,Ref:ALICEPPR}.
For this analysis, the following detector subsystems were used: the Time Projection Chamber 
(\TPC) \cite{Ref:ALICETPC}, the Silicon Pixel Detector (\SPD), two forward scintillator arrays 
(\VZERO), and two Zero Degree Calorimeters (\ZDC) \cite{Ref:ALICE}. 

We analysed a sample of about 13~million minimum--bias trigger events of Pb--Pb 
collisions at $\sqrt{s_{NN}} = 2.76$~TeV collected with the ALICE detector. The standard \mbox{ALICE} offline event 
selection criteria \cite{Ref:AliceFlow} were applied, including a collision vertex cut of $\pm 7$~cm 
along the beam axis. The collision centrality is estimated from the amplitude measured 
by the VZERO detectors \cite{Ref:ALICE}. The data 
sample is divided into centrality classes which span 0-70$\%$ of the hadronic interaction 
cross section, with the 0-5$\%$ class corresponding to the most central (i.e.~smaller impact parameter) collisions. Charged 
particles reconstructed by the \TPC~are accepted for analysis within $|\eta| < 0.8$ and 
$0.2 < p_{\rm{T}} < 5.0$~GeV/$c$. A set of requirements described in \cite{Ref:AliceFlow} were applied in order to ensure the quality of the tracks but also to reduce the contamination from secondary particles. 

\begin{figure}[!]
\begin{center}
\includegraphics[width=\linewidth]{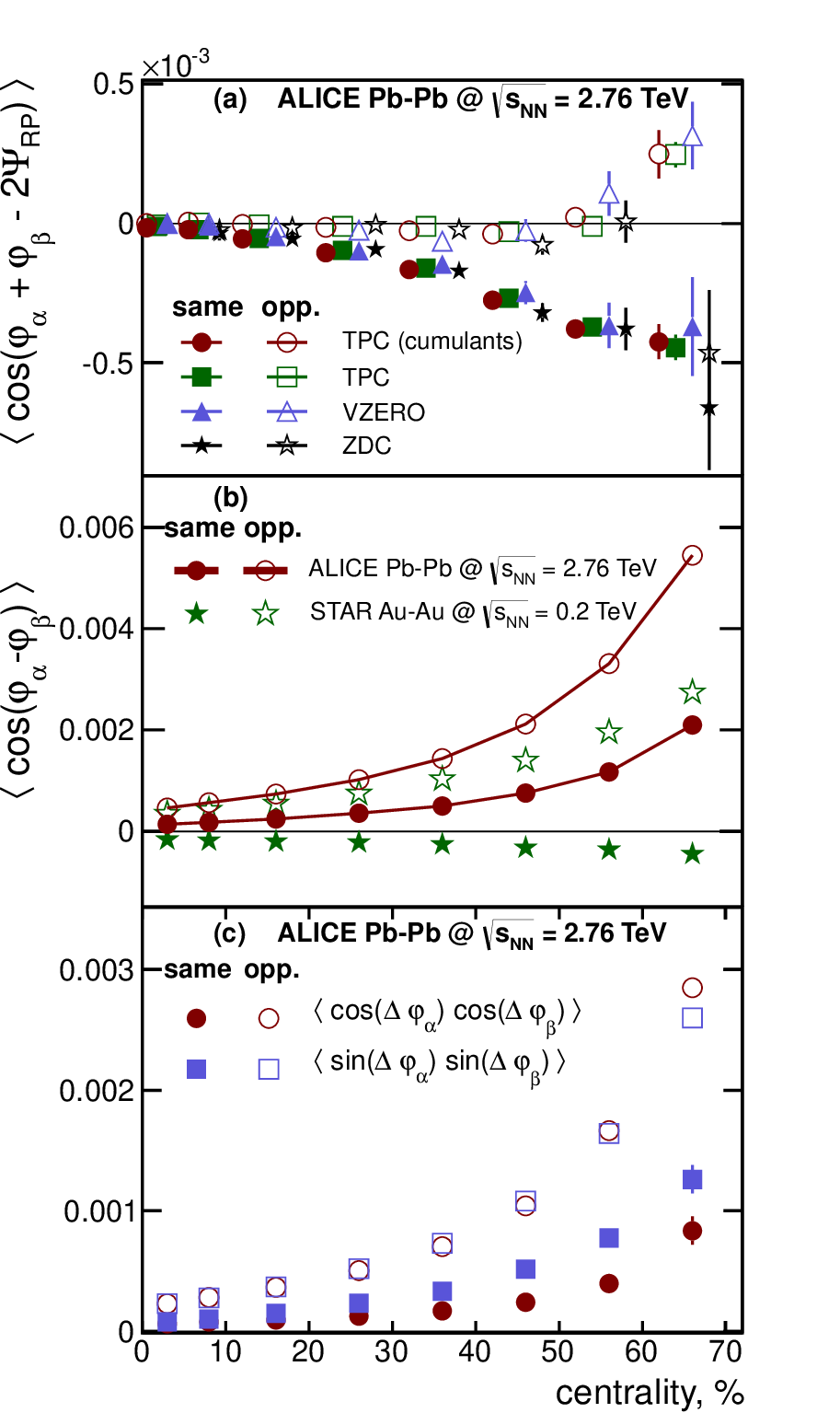}
\caption{(Colour online) (a) Centrality dependence of the correlator defined in Eq.~\ref{Eq:3ParticleCorrelator} 
measured with the cumulant method, and from correlations with the reaction plane estimated using 
the \TPC, the \ZDC~and the \VZERO~detectors. Only statistical errors are shown. The points are displaced 
slightly in the horizontal direction for visibility. (b) Centrality dependence of the two--particle correlator
defined in Eq.~\ref{Eq:2ParticleCorrelator} compared to the STAR data~\cite{Ref:STAR}. 
The width of the solid red lines indicates the systematic uncertainty of the ALICE measurement.
(c) Decomposition of the correlators into $\mean{\cos\Delta \phia\, \cos\Delta \phib}$ and 
$\mean{\sin\Delta \phia\,\sin\Delta \phib}$ terms. The ALICE results in (b) and (c) are obtained with 
the cumulant method.}
\label{fig:results}
\end{center}
\end{figure}

To evaluate the systematic uncertainties in the \mbox{analysis}, events recorded with two different magnetic field 
polarities were analysed leading to  an uncertainty below 7$\%$ for all centrality classes. The cut on the 
collision vertex was varied from $\pm$7~cm to $\pm$10~cm from the nominal collision point, with steps of 
1~cm, contributing a maximum of 5$\%$ to the total uncertainty. A bias due to the centrality determination 
was studied by using multiplicities measured by the \TPC~or the \SPD, rather than the \VZERO, and was 
found to be less than 10$\%$. Contamination due to secondary tracks that do not originate from the collision 
vertex was reduced by requiring that the distance of closest approach between tracks and the primary vertex 
is less than 2~cm. The effect of secondary tracks on the measurement was estimated by varying the cut from 
2 to 4~cm in steps of 0.5 cm, and was calculated to be below 15$\%$.  Effects due to non--uniform acceptance 
of the \TPC~were estimated to be below 2\%, and are corrected for in the analysis. A significant contribution to 
the systematic error is coming from the uncertainty in the v$_2$ measurement which is used as an estimate of 
the reaction plane resolution. The v$_2$ estimate is obtained from the $2$- and $4$--particle cumulant analysis 
\cite{Ref:AliceFlow}, which are affected in different ways by non--flow effects and flow fluctuations. For this 
analysis, v$_{2}$ was taken as the average of the two values, with half of the difference between v$_2\{2\}$ and 
v$_2\{4\}$ being attributed as the systematic uncertainty. The values of this uncertainty range from 9$\%$ for the 
20--30$\%$ centrality to 18$\%$ (24$\%$) for the 50--60$\%$ (60--70$\%$) centrality class. The differences in 
the results from the four independent analysis methods (described below) were also considered as part of the 
systematic uncertainty and were estimated to be 3$\%$ for the 20--30$\%$ and the 50--60$\%$ centrality bins 
and 47$\%$ for the most peripheral centrality class. The contributions from all effects were added in quadrature 
to calculate the total systematic uncertainty. For the correlation between pairs of particles with the same charge 
it varies from 19$\%$ (28$\%$) for the 20--30$\%$ (50--60$\%$) centrality up to 55$\%$ for the 60--70$\%$ centrality 
class. The correlations between opposite charged particles for 0--60$\%$  centrality and for the same charge pairs 
for 0--20$\%$ centrality are compatible with zero with a systematic error below $5.5\times10^{-5}$.


Figure~\ref{fig:results}a presents the centrality dependence of the three--particle correlator, defined 
in Eq.~\ref{Eq:3ParticleCorrelator}. The correlations of the same charge pairs for the positive--positive 
and negative--negative combinations are found to be consistent within statistical uncertainties and 
are combined into one set of points, labelled {\it same}. The difference between the correlations of 
pairs with same and opposite charge indicates a charge dependence with respect to the reaction 
plane, as may be expected for the CME. To test the bias from the reaction plane reconstruction, four 
independent analyses were performed. The first analysis uses a cumulant technique~\cite{Ref:Qumulants}, 
whereas for the three other analyses the orientation of the collision symmetry plane is estimated 
from the azimuthal distribution of charged particles in the \TPC, and hits in the forward \VZERO~and 
\ZDC~detectors \cite{Ref:EventPlane}. There is a very good agreement between the results obtained with the 
event plane estimated from different detectors covering a wide range in pseudo--rapidity. This allows to conclude 
that background sources due to correlations not related to the orientation of the reaction plane are negligible, 
with maybe the exception of the peripheral collisions for the pairs of particles with opposite charge.

Figure \ref{fig:results}b shows the centrality dependence of the two--particle correlator 
$\langle \cos(\varphi_{\alpha} - \varphi_{\beta}) \rangle$, as defined in Eq.~\ref{Eq:2ParticleCorrelator}, 
which helps to constrain experimentally the $P$--even background correlations. The statistical uncertainty 
is smaller than the symbol size. The two--particle correlations for the same and opposite charge combinations 
are always positive and exhibit qualitatively similar centrality dependence, while the magnitude of the 
correlation is smaller for the same charged pairs. Our results differ from those reported by the STAR 
Collaboration for Au-Au collisions at $\sqrt{s_{NN}} = 200$~GeV~\cite{Ref:STAR} for which negative 
correlations are observed for the same charged pairs.

Figure~\ref{fig:results}c shows the $\mean{\cos\Delta \phia\, \cos\Delta \phib}$ and 
$\mean{\sin\Delta \phia\,\sin\Delta \phib}$ terms separately.~For pairs of particles of the same charge, 
we observe that the $\mean{\sin\Delta \phia\,\sin\Delta \phib}$ correlations are larger than the 
$\mean{\cos\Delta \phia\,\cos\Delta \phib}$ ones. On the other hand, for pairs of opposite charge, the 
two terms are very close except for the most peripheral collisions. Further interpretation of the results 
presented in Fig.~\ref{fig:results}c in terms of in-- and out--of--plane correlations is complicated due to 
the significant non--flow contribution in $\langle \cos(\varphi_{\alpha} - \varphi_{\beta}) \rangle$.

\begin{figure}[t]
\includegraphics[width=\linewidth]{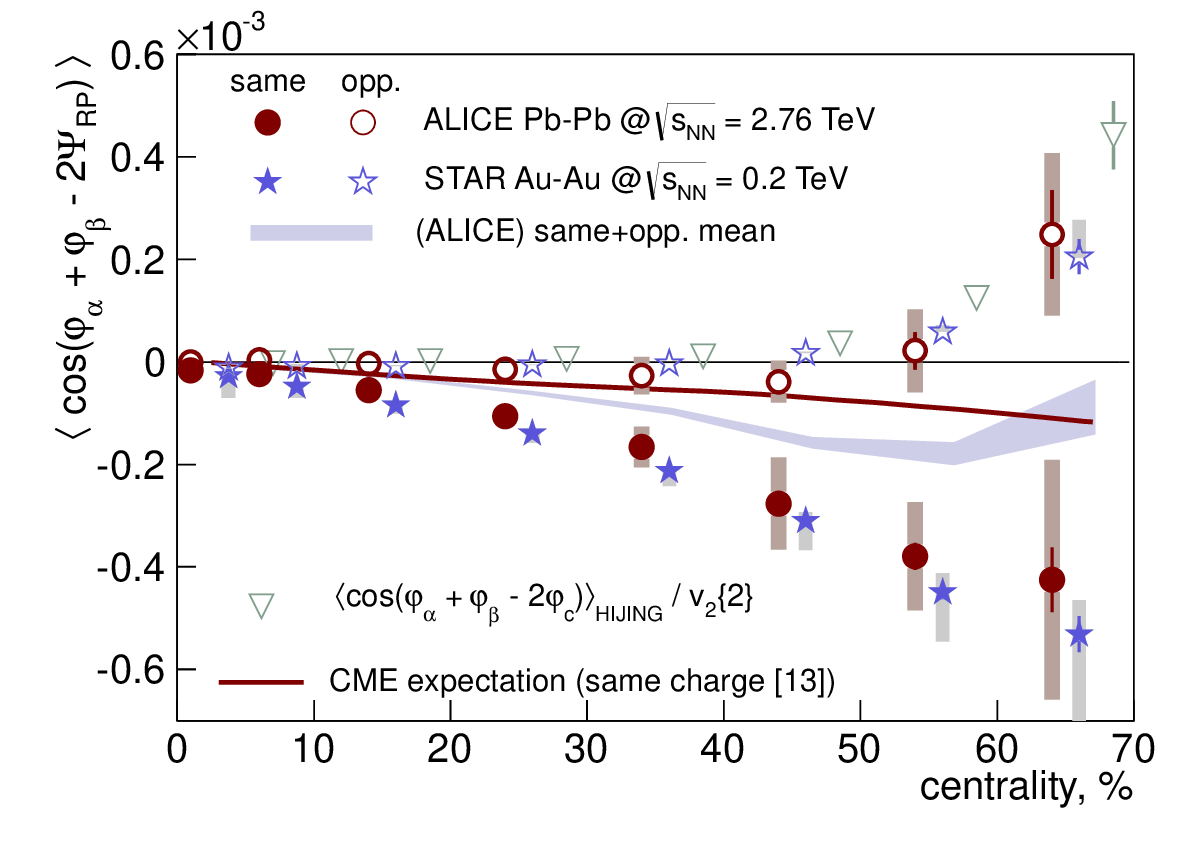}
\caption{(Colour online) The centrality dependence of the three--particle correlator defined in 
Eq.~\ref{Eq:3ParticleCorrelator}. The circles indicate the ALICE results obtained from the 
cumulant analysis. The stars show the STAR data from \cite{Ref:STAR}. The triangles 
represent the genuine three--particle correlations ($\langle \cos(\varphi_{\alpha} + \varphi_{\beta} -2\varphi_c) \rangle$) from HIJING \cite{Ref:Hijing} corrected for the experimentally measured v$_2\{2\}$ \cite{Ref:AliceFlow}. Points are displaced horizontally for visibility.
A model prediction for the same sign correlations incorporating the \cme~for LHC energies~\cite{Ref:Toneev} 
is shown by the solid line. The shaded band represents the centrality dependence of the charge independent correlations.}
\label{fig:comparisonRP}
\end{figure}

\begin{figure*}[t]
\includegraphics[width=0.9\linewidth]{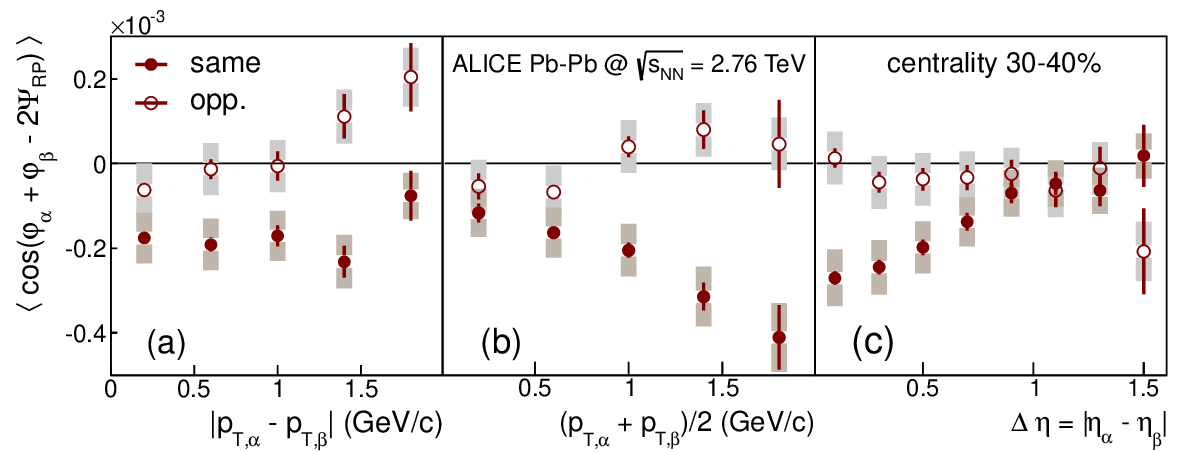}
\caption{(Colour online) The three--particle correlator defined in Eq.~\ref{Eq:3ParticleCorrelator}
as a function of (a) the transverse momentum difference, $|p_{{\rm t},\alpha}-p_{{\rm t},\beta}|$, (b) 
the average transverse momentum, $(p_{{\rm t},\alpha}+p_{{\rm t},\beta})/2$, and (c) the rapidity 
separation, $|\eta_{\alpha}-\eta_{\beta}|$, of the charged particle pair of same (closed symbols) 
and opposite (open symbols) sign.}
\label{fig:3pDifferential}
\end{figure*}

Figure~\ref{fig:comparisonRP} presents the three--particle correlator $\langle \cos(\varphi_{\alpha} + \varphi_{\beta} - 2\Psi_{\rm RP}) \rangle$ as a function of the collision centrality compared to model calculations and 
results for RHIC energies. The statistical uncertainties are represented by the error bars. The shaded 
area around the points indicates the systematic uncertainty based on the different sources described 
above. Also shown in Fig.~\ref{fig:comparisonRP} are STAR results~\cite{Ref:STAR}. The small difference 
between the LHC and the RHIC data indicates little or no energy dependence for the three--particle 
correlator when changing from the collision energy of $\sqrt{s_{NN}} = 0.2$ TeV to 2.76~TeV.

In Fig.~\ref{fig:comparisonRP}, the ALICE data are compared to the expectations from the HIJING model \cite{Ref:Hijing}. 
The HIJING results for the three--particle correlations are divided by the experimentally measured value of v$_2$ (i.e. 
$\langle \cos(\varphi_{\alpha} + \varphi_{\beta} -2\varphi_c) \rangle$/v$_2\{2\}$) as reported in \cite{Ref:AliceFlow} due 
to the absence of collective azimuthal anisotropy in this model. Since the points do not exhibit any significant difference 
between the correlations of pairs with same and opposite charge, they were averaged in the figure. The correlations from 
HIJING show a significant increase in the magnitude for very peripheral collisions. This can be attributed to correlations 
not related to the reaction plane orientation, in particular, from jets \cite{Ref:STAR}. 

The results from ALICE in Fig.~\ref{fig:comparisonRP} show a strong correlation of pairs with the same charge and 
simultaneously a very weak correlation for the pairs of opposite charge. This difference in the correlation magnitude 
depending on the charge combination could be interpreted as ``quenching'' of the charge correlations for the case 
when one of the particles is emitted toward the centre of the dense medium created in a heavy--ion collision 
\cite{Ref:BField1,Ref:TheoryExpectations}. An alternative explanation can be provided by a recent suggestion 
\cite{Ref:TeaneyYan} that the value of the charge independent version of the correlator defined in 
Eq.~\ref{Eq:3ParticleCorrelator} is dominated by directed flow fluctuations. The sign and the magnitude 
of these fluctuations based on a hydrodynamical model calculation for RHIC energies~\cite{Ref:TeaneyYan} 
appear to be very close to the measurement. Our results for charge independent correlations are given 
by the shaded band in Fig.~\ref{fig:comparisonRP}.

The thick solid line in Fig.~\ref{fig:comparisonRP} shows a prediction \cite{Ref:Toneev} for the same 
sign correlations due to the CME at LHC energies. The model makes no prediction of the absolute 
magnitude of the effect, and can only describe the energy dependence by taking into account the 
duration and time evolution of the magnetic field. It predicts a decrease of correlations by about 
a factor of five from RHIC to LHC, which would significantly underestimate the observed magnitude 
of the same sign correlations seen at  the LHC. At the same time in \cite{Ref:TheoryExpectations,Ref:Zhitnitsky},
it was suggested  that the CME might have the same magnitude at the LHC and at RHIC energies.

Figure~\ref{fig:3pDifferential} shows the dependence of the three--particle correlator on the transverse 
momentum difference, $|p_{{\rm T},\alpha}-p_{{\rm T},\beta}|$, the average transverse momentum, 
$(p_{{\rm T},\alpha}+p_{{\rm T},\beta})/2$, and the pseudo--rapidity separation, $|\eta_{\alpha}-\eta_{\beta}|$,
of the pair for the 30--40$\%$ centrality range. The pairs of opposite charge do not show any significant 
dependence on the pseudo--rapidity difference, while there is a dependence on $|p_{{\rm T},\alpha}-p_{{\rm T},\beta}|$ (stronger) and $(p_{{\rm T},\alpha}+p_{{\rm T},\beta})/2$ (weaker). The correlations for pairs of particles of the same charge show no strong dependence on the \pT difference, allowing to exclude any type of short range correlations (e.g. HBT) as the main source of the effect. In addition, it is seen that the magnitude of the same charge correlations increases with increasing average \pT of the pair. This observation is in contradiction with the initial expectation from theory \cite{Ref:TheoryExpectations} that the effect should originate from low \pT particles. The  dependence of the correlations on the $|\eta_{\alpha}-\eta_{\beta}|$ indicates a width of one unit in pseudo--rapidity, beyond which the value of $\langle \cos(\varphi_{\alpha} + \varphi_{\beta} - 2\Psi_{\rm RP}) \rangle$ is close to zero up to $\Delta \eta \approx 1.5$. Similar results were reported also at RHIC energies \cite{Ref:STAR}. At the moment there are no quantitative model calculations of the charge dependent differential correlations.


In summary, we have measured the charge dependent azimuthal correlations in Pb--Pb collisions at 
$\sqrt{s_{NN}} = 2.76$~TeV at the LHC using the ALICE detector. Both two-- and three--particle correlations 
are reported. A clear difference in the correlation strength between the same and opposite charge particle 
combinations is observed. The centrality dependence of these correlations is in qualitative agreement with 
a charge--dependent separation relative to the reaction plane. Our results are not described by the only 
available quantitative model prediction of the CME for the LHC energy. The lack of realistic model calculations 
for the centrality and pair differential dependencies based on models incorporating CME and possible background 
contributions does not allow to make a firm conclusion regarding the nature of the charge dependent correlations 
originally observed at RHIC and now established at the LHC. The observation of a small collision energy 
dependence of the three-particle correlation and the simultaneous significant change in the two-particle correlations 
between top RHIC and LHC energies put stringent constraints on models built to interpret such correlations.
Analyses of higher harmonic correlations are planned and may yield a better understanding of the complex charge 
dependent correlations seen at LHC energies.

%% file: acknowledgements.tex
The ALICE collaboration would like to thank all its engineers and technicians for their invaluable contributions to the construction of the experiment and the CERN accelerator teams for the outstanding performance of the LHC complex.
\\
The ALICE collaboration acknowledges the following funding agencies for their support in building and
running the ALICE detector:
 \\
Calouste Gulbenkian Foundation from Lisbon and Swiss Fonds Kidagan, Armenia;
 \\
Conselho Nacional de Desenvolvimento Cient\'{\i}fico e Tecnol\'{o}gico (CNPq), Financiadora de Estudos e Projetos (FINEP),
Funda\c{c}\~{a}o de Amparo \`{a} Pesquisa do Estado de S\~{a}o Paulo (FAPESP);
 \\
National Natural Science Foundation of China (NSFC), the Chinese Ministry of Education (CMOE)
and the Ministry of Science and Technology of China (MSTC);
 \\
Ministry of Education and Youth of the Czech Republic;
 \\
Danish Natural Science Research Council, the Carlsberg Foundation and the Danish National Research Foundation;
 \\
The European Research Council under the European Community's Seventh Framework Programme;
 \\
Helsinki Institute of Physics and the Academy of Finland;
 \\
French CNRS-IN2P3, the `Region Pays de Loire', `Region Alsace', `Region Auvergne' and CEA, France;
 \\
German BMBF and the Helmholtz Association;
\\
General Secretariat for Research and Technology, Ministry of
Development, Greece;
\\
Hungarian OTKA and National Office for Research and Technology (NKTH);
 \\
Department of Atomic Energy and Department of Science and Technology of the Government of India;
 \\
Istituto Nazionale di Fisica Nucleare (INFN) of Italy;
 \\
MEXT Grant-in-Aid for Specially Promoted Research, Ja\-pan;
 \\
Joint Institute for Nuclear Research, Dubna;
 \\
National Research Foundation of Korea (NRF);
 \\
CONACYT, DGAPA, M\'{e}xico, ALFA-EC and the HELEN Program (High-Energy physics Latin-American--European Network);
 \\
Stichting voor Fundamenteel Onderzoek der Materie (FOM) and the Nederlandse Organisatie voor Wetenschappelijk Onderzoek (NWO), Netherlands;
 \\
Research Council of Norway (NFR);
 \\
Polish Ministry of Science and Higher Education;
 \\
National Authority for Scientific Research - NASR (Autoritatea Na\c{t}ional\u{a} pentru Cercetare \c{S}tiin\c{t}ific\u{a} - ANCS);
 \\
Federal Agency of Science of the Ministry of Education and Science of Russian Federation, International Science and
Technology Center, Russian Academy of Sciences, Russian Federal Agency of Atomic Energy, Russian Federal Agency for Science and Innovations and CERN-INTAS;
 \\
Ministry of Education of Slovakia;
 \\
Department of Science and Technology, South Africa;
 \\
CIEMAT, EELA, Ministerio de Educaci\'{o}n y Ciencia of Spain, Xunta de Galicia (Conseller\'{\i}a de Educaci\'{o}n),
CEA\-DEN, Cubaenerg\'{\i}a, Cuba, and IAEA (International Atomic Energy Agency);
 \\
Swedish Research Council (VR) and Knut $\&$ Alice Wallenberg
Foundation (KAW);
 \\
Ukraine Ministry of Education and Science;
 \\
United Kingdom Science and Technology Facilities Council (STFC);
 \\
The United States Department of Energy, the United States National
Science Foundation, the State of Texas, and the State of Ohio.

%% file: references.tex

%% file: parityPaper.bbl
\begin{thebibliography}{99}

\bibitem{Ref:TDLee}{T.~D.~Lee, Phys. Rev. \textbf{D8}, 1226 (1973). 
T.~D.~Lee and G.~C.~Wick, Phys. Rev. \textbf{D9}, 2291 (1974).}

\bibitem{Ref:Morley}{P.~D.~Morley and I.~A.~Schmidt, Z. Phys. \textbf{C26}, 627 (1985).}

\bibitem{Ref:Kharzeev}{D.~Kharzeev, R.~D.~Pisarski and M.~H.~G.~Tytgat, Phys. Rev. Lett. \textbf{81}, 512 (1998). 
D.~Kharzeev and R.~D.~Pisarski, Phys. Rev. \textbf{D61}, 111901 (2000).}

\bibitem{Ref:InstantonsSphalerons} {
T.~Sch\"{a}fer and E.~V.~Shuryak, Rev. Mod. Phys. \textbf{70}, 323 (1998).
}

\bibitem{Ref:ChernSimons} { S.~S.~Chern, J.~Simons, Ann. Math. \textbf{99}, 48 (1974).}

\bibitem{Ref:BField1}{D.~Kharzeev, Phys. Lett. \textbf{B633}, 260 (2006). 
K.~Fukushima, D.~E.~Kharzeev and H.~J.Warringa, Phys. Rev. \textbf{D78}, 074033 (2008).}

\bibitem{Ref:TheoryExpectations} {
D.~Kharzeev, L.~D.~McLerran and H.~J.~Warringa, Nucl. Phys. \textbf{A803}, 227 (2008).}

\bibitem{Ref:STAR}{B.~I.~Abelev \textit{et al.} [STAR Collaboration], Phys. Rev. Lett. \textbf{103}, 251601 (2009). 
B.~I.~Abelev \textit{et al.} [STAR Collaboration], Phys. Rev. \textbf{C81}, 54908 (2010).}

\bibitem{Ref:Muller} 
  B.~M\"{u}ller and A.~Sch\"{a}fer,
  Phys.\ Rev.\ C\ {\bf 82}, 057902  (2010)
  [arXiv:1009.1053 [hep-ph]].

\bibitem{Ref:Pratt} 
  S.~Schlichting and S.~Pratt,
  Phys.\ Rev.\ C\ {\bf 83}, 014913  (2011). 
  S.~Pratt, S.~Schlichting and S.~Gavin,
  Phys.\ Rev.\ C\ {\bf 84}, 024909  (2011)

\bibitem{Ref:Koch}{A.~Bzdak, V.~Koch and J.~Liao, Phys. Rev. \textbf{C81}, 031901  (2010). 
A.~Bzdak, V.~Koch and J.~Liao, Phys. Rev. \textbf{C83}, 014905 (2011).}

\bibitem{Ref:Zhitnitsky} {A.~R.~Zhitnitsky, Nucl. Phys. \textbf{A853}, 135 (2011). 
A.~R.~Zhitnitsky, Nucl. Phys. \textbf{A886}, 18 (2012).}

\bibitem{Ref:Toneev} {V.~D.~Toneev and V.~Voronyuk, arXiv:1012.1508 [nucl-th].}

\bibitem{Ref:FlowHarmonics} {
S.~Voloshin and Y.~Zhang, Z. Phys. \textbf{C70}, 665 (1996). 
A.~M.~Poskanzer and S.~Voloshin, Phys. Rev. \textbf{C58}, 1671 (1998).}

\bibitem{Ref:Sergey3particleCorrelator}{S.~A.~Voloshin, Phys. Rev. \textbf{C70}, 057901 (2004).}

\bibitem{Ref:TeaneyYan}{D.~Teaney and L.~Yan, Phys. Rev. \textbf{C83}, 064904 (2011).}

\bibitem{Ref:ALICE} {K.~Aamodt \textit{et al.} [ALICE Collaboration], JINST \textbf{3}, S08002 (2008).}

\bibitem{Ref:ALICEPPR} {K.~Aamodt \textit{et al.} [ALICE Collaboration], J. Phys. \textbf{G30}, 1517 (2004). 
K.~Aamodt \textit{et al.} [ALICE Collaboration], J. Phys. \textbf{G32}, 1295 (2006).}

\bibitem{Ref:ALICETPC} {J.~Alme \textit{et al.} [ALICE Collaboration], Nucl. Instrum. Meth. \textbf{A622}, 316 (2010).}

\bibitem{Ref:AliceFlow}
{K.~Aamodt \textit{et al.} [ALICE Collaboration], Phys. Rev. Lett. \textbf{105}, 252302 (2010).}

\bibitem{Ref:Qumulants}{A.~Bilandzic, R.~Snellings and S.~Voloshin, Phys. Rev. \textbf{C83}, 044913 (2011).}

\bibitem{Ref:EventPlane} {S.~Voloshin, A.~M.~Poskanzer, and R.~Snellings, in \emph{Relativistic Heavy Ion Physics}, Landolt-Bornstein Vol. 1 (Springer-Verlag, Berlin, 2010), pp. 5Ð54.}

\bibitem{Ref:Hijing}
{M.~Gyulassy and X.~N.~Wang, Comput. Phys. Commun. \textbf{83}, 307 (1994). 
X.~N.~Wang and M.~Gyulassy, Phys. Rev. \textbf{D44}, 3501 (1991).}







\end{thebibliography}
